\documentclass[a4paper,fleqn,usenatbib]{mnras}

\usepackage[caption=false]{subfig}

\usepackage[T1]{fontenc}
\usepackage{ae,aecompl}

\usepackage{graphicx}	
\usepackage{amsmath}	
\usepackage{amssymb}	
\usepackage{float}

\def\aj{AJ}%
\def\apj{ApJ}%
\def\apjl{ApJ}%
\def\apjs{ApJS}%
\def\aap{A\&A}%
\def\aaps{A\&AS}%
\def\icarus{Icarus}%
\def\mnras{MNRAS}%
\def\nat{Nature}%
\title[ETVs, EDVs, and RVs can reveal S-type planets]{How eclipse time variations, eclipse duration variations, and radial velocities can reveal S-type planets in close eclipsing binaries}
\author[M. Oshagh, R. Heller, and S. Dreizler]{
M. Oshagh$^{1}$\thanks{E-mail: moshagh@astro.physik.uni-goettingen.de}
, R. Heller$^{2}$\thanks{E-mail: heller@mps.mpg.de}
, and S. Dreizler$^{1}$
\\
$^{1}$Institut f\"ur Astrophysik, Georg-August-Universit\"at,
Friedrich-Hund-Platz 1, 37077 G\"ottingen, Germany\\
$^{2}$Max Planck Institute for Solar System Research, Justus-von-Liebig-Weg 3, 37077 G\"ottingen, Germany
}

\date{Accepted XXX. Received YYY; in original form ZZZ}

\pubyear{2016}

\begin{document}
\label{firstpage}
\pagerange{\pageref{firstpage}--\pageref{lastpage}}
\maketitle

\begin{abstract}
While about a dozen transiting planets have been found in wide orbits around an inner, close stellar binary (so-called ``P-type planets''), no planet has yet been detected orbiting only one star (a so-called ``S-type planet'') in an eclipsing binary. This is despite a large number of eclipsing binary systems discovered with the \textit{Kepler} telescope. Here we propose a new detection method for these S-type planets, which uses a correlation between the stellar radial velocities (RVs), eclipse timing variations (ETVs), and eclipse duration variations (EDVs). We test the capability of this technique by simulating a realistic benchmark system and demonstrate its detectability with existing high-accuracy RV and photometry instruments. We illustrate that, with a small number of RV observations, the RV-ETV diagrams allows us to distinguish between prograde and retrograde planetary orbits and also the planetary mass can be estimated if the stellar cross-correlation functions can be disentangled. We also identify a new (though minimal) contribution of S-type planets to the Rossiter-McLaughlin effect in eclipsing stellar binaries. We finally explore possible detection of exomoons around transiting luminous giant planets and find that the precision required to detect moons in the RV curves of their host planets is of the order of ${\rm cm\,s}^{-1}$ and therefore not accessible with current instruments.
\end{abstract}



\section{Introduction}

Among the detection of thousands of extrasolar planets and candidates around single stars, the \textit{Kepler} telescope has also delivered the first transit observations of planets in stellar multiple systems, eleven in total as of today \citep{2011Sci...333.1602D,2012Natur.481..475W,2012Sci...337.1511O,2012ApJ...758...87O,2013ApJ...768..127S,2014ApJ...784...14K,2015ApJ...809...26W,2016ApJ...827...86K}. Most of these binaries show mutual eclipses, therefore allowing precise radius estimates of both stars and planets in a given system. All of these planets are classical ``circumbinary planets'', or P-type planets \citep{1982OAWMN.191..423D}, orbiting on the circumbinary orbit around an inner stellar binary (or their common center of mass) on a wide orbit. In other systems, the planetary transit cannot be observed directly, but stellar eclipse timing variations (ETVs) still signify the presence of one or more P-type planets \citep{Beuermann-10,2010ApJ...708L..66Q,Beuermann-11,Schwarz-11,2012MNRAS.422L..24Q,Beuermann-12,Beuermann-13,Marsh-14,2016MNRAS.460.3598S}, some of which are still being disputed \citep{2012MNRAS.425..749H,2013MNRAS.435.2033H, Bours-16}.

Planets also exist in S-type configurations (``S'' for satellite), where the planet orbits a star in a close orbit, while both the planet and its host star are gravitationally bound to an additional star.\footnote{A third configuration is referred to as `L-type'' (or sometimes ``T-type'', ``T'' for Trojan), where the planet is located at either the $L_4$ or $L_5$ Lagrangian point \citep{1986A&A...167..379D}. Such planets have been suggested, and could possibly be detectable via eclipse time variations \citep{1995EM&P...71..153S,2015MNRAS.453.2308S}, but remain undiscovered as of today.}  Many of the known exoplanets are indeed S-type planets, mostly in wide-orbit binaries. A handful of S-type planets are orbiting stellar binaries with separations $\lesssim20$\,AU (the tightest known binary which hosts a s-type planet is KOI-1257 system with the semi-major axis of 5.3 AU \citep{Santerne-14}) , but none of these systems are eclipsing. In other words, there has been no detection of an S-type planet in a stellar eclipsing binary.

Formation of S-type planets, particularly of giant planets, in close binaries may be hampered by gravitational and dynamical perturbations from the stars \citep{Whitmire-98, Nelson-00,Thebault-04, Haghighipour-09, Thebault-14}. However, to explain the discovered S-type planets in close (non-eclipsing) binaries, several solutions have been proposed. For instance, a binary orbit might be initially wide and tighten only after planet formation and migration have completed \citep{Malmberg-07,Thebault-09, Thebault-14}. Alternatively, S-type planets, giant or terrestrial, in close eclipsing binaries could build up through other formation scenarios, such as gravitational instabilities \citep{Boss-97} or in-situ core-accretion \citep{Boss-06, Mayer-07, Duchene-10}.

\begin{figure}
\includegraphics[width=1.0\linewidth]{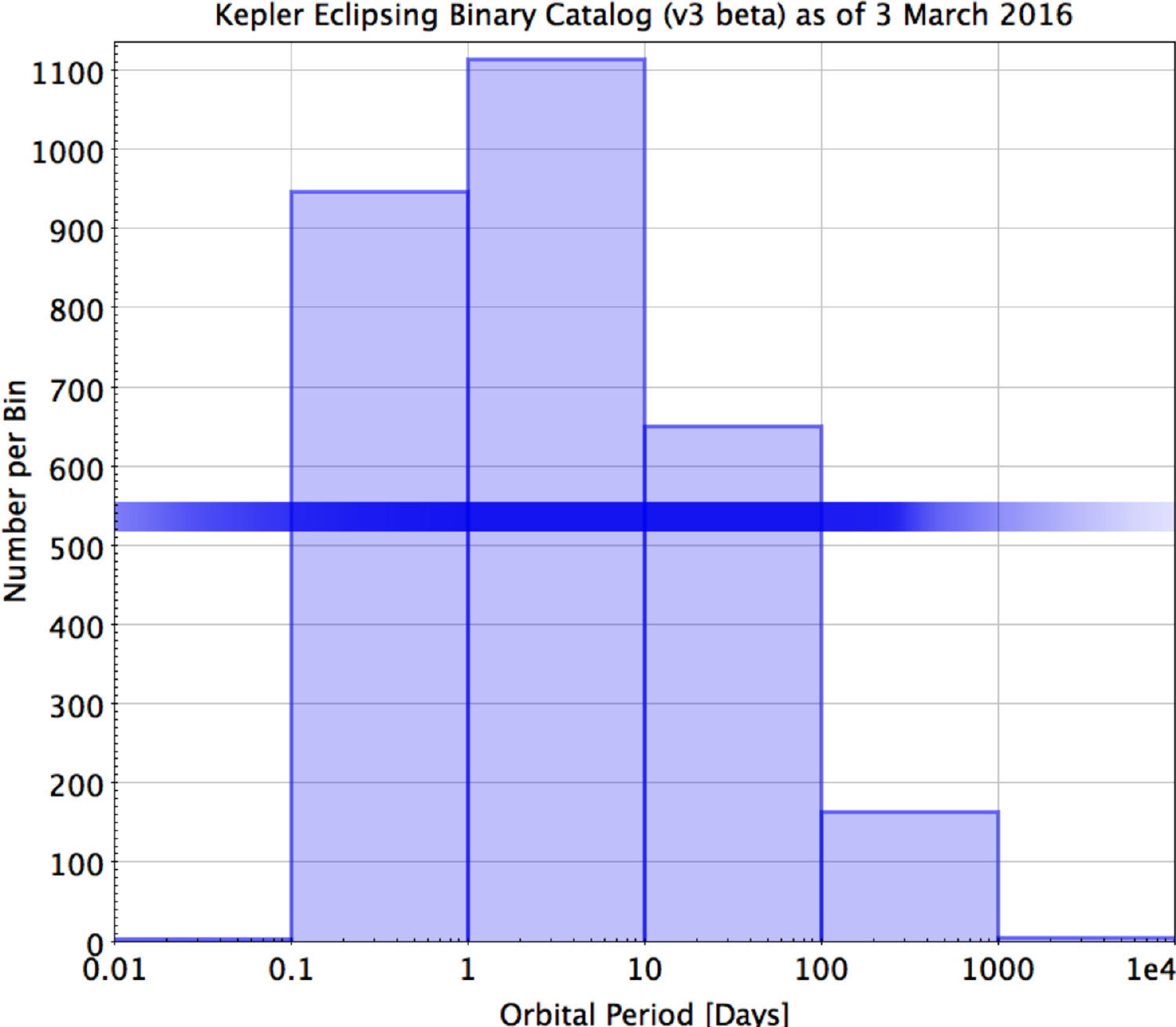}
 \caption{Period distribution of almost 3,000 stellar eclipsing binaries observed by \textit{Kepler} space telescope from \href{http://keplerebs.villanova.edu}{http://keplerebs.villanova.edu} \citep{2016AJ....151...68K}. The horizontal bar represents a density distribution of the histogram with dark
blue corresponding to the highest densities and white corresponding to zero.}
  \label{fig:periods}
\end{figure}

Therefore, S-type planets in close eclipsing binaries might be expected to be rare from both a formation and an orbital stability point of view \citep{1986A&A...167..379D,1999AJ....117..621H,2012MNRAS.425..749H}. Yet, the discoveries of planets around pulsars \citep{1992Natur.355..145W}, of the previously unpredicted families of hot Jupiters \citep{1995Natur.378..355M}, and of circumbinary planets \citep{2011Sci...333.1602D} remind us that formation theories do not necessarily deliver ultimate constraints on the actual presence of planet populations. As a consequence, the discovery of an S-type planet in close eclipsing binaries would have major impact on our understanding of planet formation and evolution. The detection of almost 3,000 close eclipsing binaries with \textit{Kepler} \citep{2016AJ....151...68K} suggests that a dedicated search could indeed be successful. In Figure~\ref{fig:periods}, we show the orbital period ($P$) distribution of these systems.\footnote{\href{http://keplerebs.villanova.edu}{http://keplerebs.villanova.edu}} Note that binaries with $P<10$\,d have Hill spheres of about ten solar radii or smaller, so any planets might be subject to removal over billions of years. In systems with periods beyond $10$\,d, however, S-type planets might be long-lived and present today.

Here we report on a new method to detect S-type planets in close eclipsing binaries using radial velocities (RVs), eclipse timing variations (ETVs), and eclipse duration variation (EDVs). It is an extension of recently developed methods to find extrasolar moons around transiting exoplanets via the planet's transit timing variations \citep[TTVs;][]{1999A&AS..134..553S} and transit duration variations \citep[TDVs;][]{2009MNRAS.392..181K} independent of the moon's direct photometric signature. In comparison to S-type exoplanet searches based on ETVs only, the combination of ETVs and EDVs provides a non-degenerate solution for the planet's mass ($M_{\rm p}$) and its semi-major axis ($a$) around the secondary star \citep{2009MNRAS.392..181K}. The RV method, using proper treatment of the stellar cross-correlation functions (CCFs) as described in this paper, offers additional measurements of $M_{\rm p}$, which must be consistent with the value derived from the ETV-EDV approach. What is more, the RV-ETV diagram offers a means to determine the sense of orbital motion of the planet (prograde or retrograde).

This paper is organised as follows: In Section 2 we describe the 
basic idea behind our method and we describe the simulations which we perform to evaluate the feasibility of our method. In Section 3. we present the results of our tests. We summarise our conclusions in Section 4.

\begin{figure*}
\includegraphics[width=.8\linewidth]{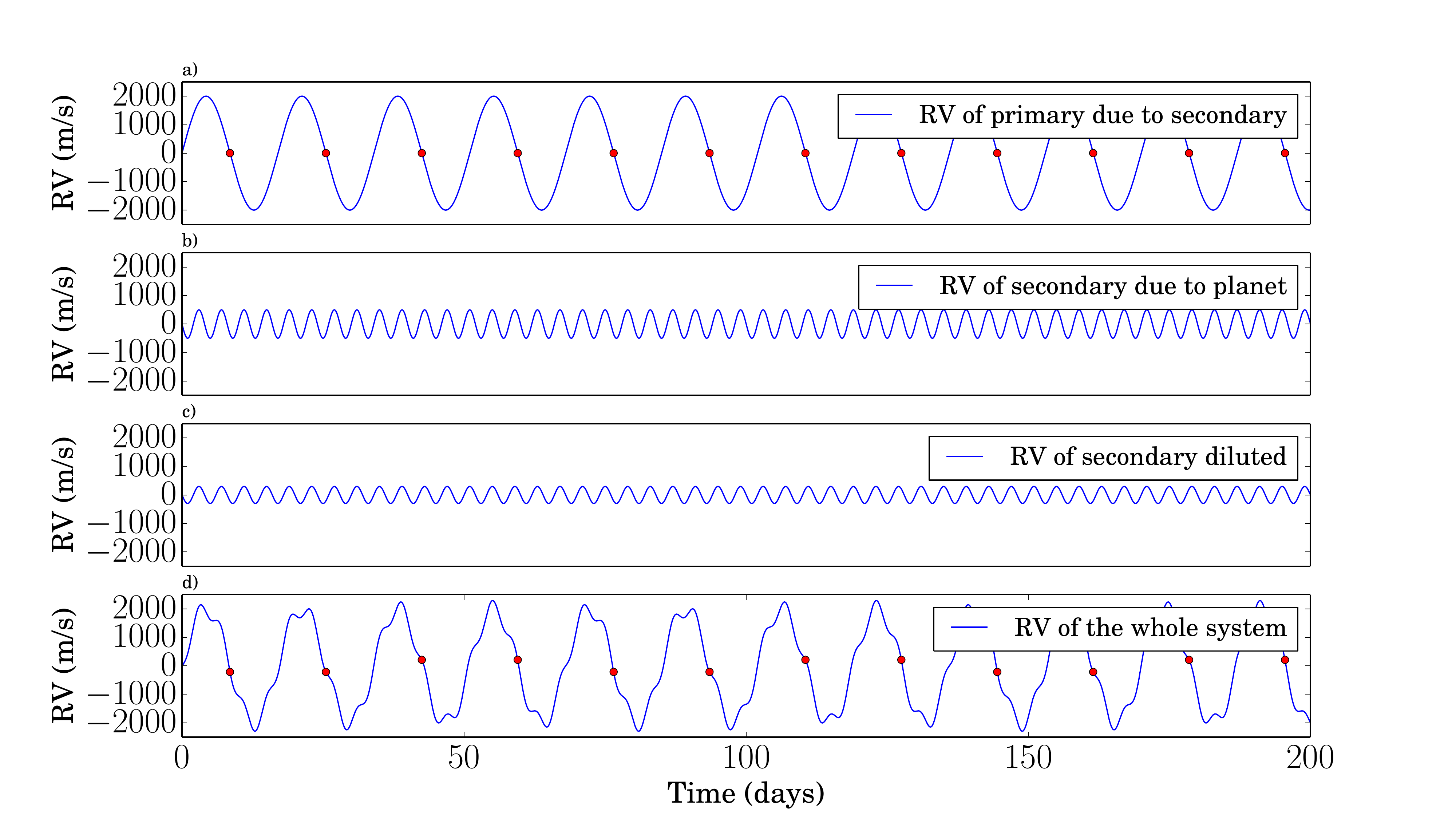}
\caption{Radial velocity components of a binary system with an S-type planet around the secondary star. Panel {\bf a)} shows the RV of the primary due to the gravitational pull of the secondary. Red dots denote times of eclipses, when the secondary star is in front of the primary (as seen from Earth) and the RV signal is zero. Panel {\bf b)} illustrates the true RV variation of the secondary star due to its S-type planet and panel {\bf c)} depicts the secondary's RV, which is diluted by the brightness contrast between the primary and the secondary. Panel {\bf d)} shows the total RV of the system, i.e. the sum of {\bf a)} and {\bf c)}.}
  \label{fig:RVconcept}
\end{figure*}

\begin{figure}
  \centering
  \includegraphics[width=1\linewidth]{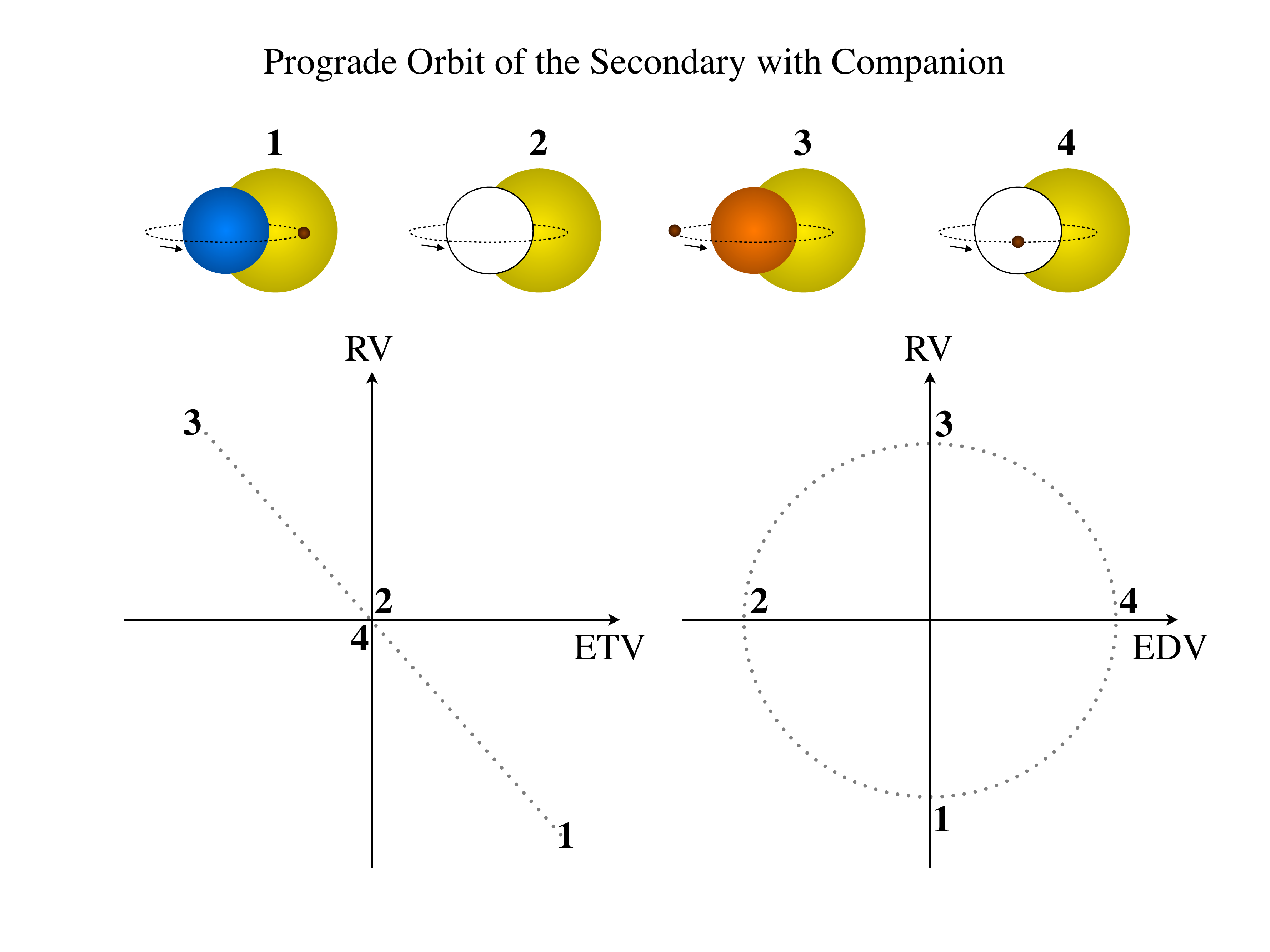}\\
  \vspace{0.2cm}
  \includegraphics[width=1\linewidth]{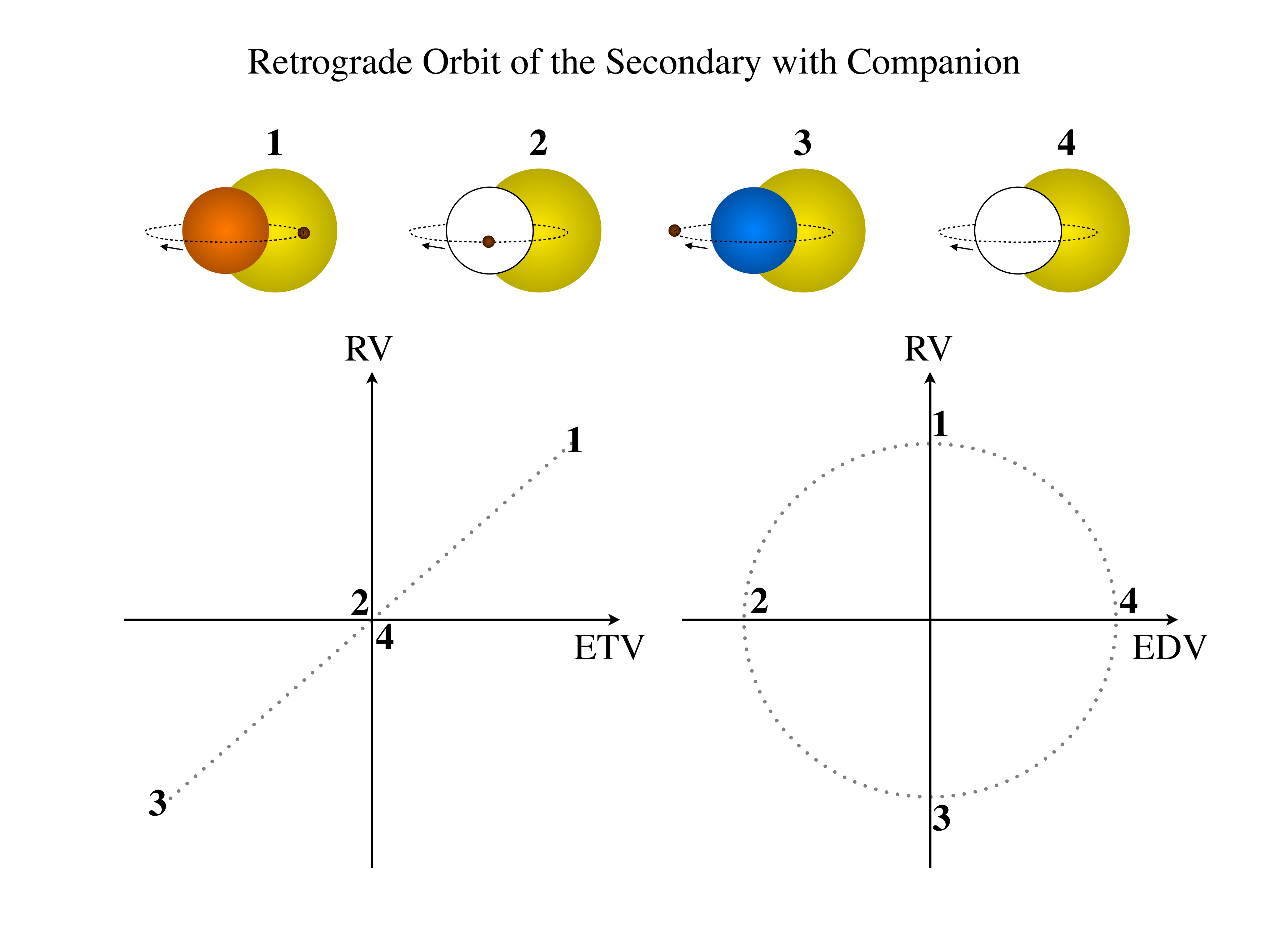}
\caption{{\bf Top}: Illustration of the RV-ETV and RV-EDV correlations for a prograde S-type planet (small black circle) around a secondary star (intermediate circle) eclipsing a primary star (big yellow circle). The colors of the secondary indicate its instantaneous RV. Blue means a negative RV-shift (toward the observer), white means zero RV, red means positive contribution (away from the observer). Arrows indicate the planet's direction of motion around the secondary. Images 1-4 depict four specific scenarios for the secondary-planet orbital geometry during eclipse, which can be found in the respective diagrams. {\bf Bottom}: Same as top but for a retrograde planet.}
\label{fig:RVETVEDV_concept}
\end{figure}

\section{Methods}

\subsection{Principles of RV, ETV, and EDV correlations}
\label{sec:principles}

Consider an eclipsing binary system consisting of a primary star, which shall be the more massive one, and a secondary star. Without loss of generality, the secondary star shall be orbited by a close-in planet. Then the planet will cause the secondary star to perform a curly, wobbly orbit around the center of mass of the stellar binary. Hence, the eclipses will not be exactly periodic: sometimes they will be too early, at other times they will be too late compared to the average eclipse period. The resulting ETVs are essentially equivalent to the previously predicted TTVs of transiting planets due to the gravitational perturbations of an extrasolar moon \citep{1999A&AS..134..553S}. Additionally, the tangential orbital velocity component of the secondary star in the star-planet system causes EDVs, which are analogous to the transit duration variations of a transiting planet with a moon \citep{2009MNRAS.392..181K}. Both ETVs and EDVs (or TTVs and TDVs) are phase-shifted by $\pi/2$, that is, they form an ellipse in the ETV-EDV (or TTV-TDV) diagram \citep{2016A&A...591A..67H}. By using high-precision photometric time series, e.g. with the \textit{Kepler} space telescope, very high-precision ETV-EDV measurements can be obtained. \textit{Kepler}'s median timing precision turned out to be $\lesssim30$\,s \citep{2013arXiv1309.1176W} and ETV accuracies obtained by \citet{2016MNRAS.455.4136B} are typically below this level.

Beyond ETVs and EDVs, the planet imposes a periodic RV variation onto its host star. RV observations of the secondary star, however, would be challenging because, first, the spectrum of the secondary star would be diluted by the primary star; and second, the RV signal of the secondary star will be on top of a much more pronounced RV signal of the stellar binary. To tackle the latter aspect, we propose to obtain the RV measurements exactly just before or after eclipse. Thus, as the orbital geometry of the stellar binary will be consistent during each such RV measurement, the RV component of the stellar binary will be equal for each measurement as well. In particular, the binary's RV component will be zero (see Figure~\ref{fig:RVconcept}a) and the observed RV will be due to the secondary's motion around the secondary-planet center of mass only (see Figure~\ref{fig:RVconcept}d). By limiting ourselves to measurements just before or after eclipse, we also avoid contributions of light travel time effects to our ETV measurements \citep{1922BAN.....1...93W,1952ApJ...116..211I}.

In Figure~\ref{fig:RVETVEDV_concept}, we illustrate the correlation of ETVs with RVs and of EDVs with RVs acquired around eclipse. The upper panel refers to a prograde planetary orbit, where the planet's sense of orbital motion is the same as the stellar binary's sense of orbital motion; the lower panel depicts a retrograde planetary orbit. The key feature of the RV-ETV correlation is in the slope of the observed curve. A negative slope (upper panel) indicates a prograde planetary orbit, whereas a positive slope (lower panel) is suggestive of a retrograde orbit. We arbitrarily normalized both the RV and the EDV amplitudes to obtain a circle. In general, however, the RV-EDV figure will be an ellipse with its semi-major and semi-minor axis given by the RV and EDV amplitudes (or vice versa).

\subsection{Feasibility tests for S-type planets and exomoons}
\label{sec:feasibility}

To assess the feasibility of our method, we simulate observations of two toy systems.

\subsubsection{A Sun-like primary and a K-dwarf secondary with an Jupiter-sized S-type planet}
\label{sec:toy1}

Our first test case consists of a Sun-like G-dwarf primary star and K-dwarf secondary star hosting an S-type Jupiter-style planet. The orbital period of the binary is chosen to be 50\,d and the planet is put in a 4\,d orbit. This choice is dominantly motivated by constraints on the long-term orbital stability of the planet. Numerical simulations by \citet{2006MNRAS.373.1227D} showed that the outermost stable orbit of a prograde satellite is at about half the Hill radius around the secondary, which translates into a maximum orbital period for the planet of about 1/9 the orbital period of the binary \citep{2009MNRAS.392..181K}.

We consider the CCFs of the publicly available HARPS spectra of 51\,Peg (our primary) and HD\,40307 (our secondary) as reference stars.\footnote{\href{http://archive.eso.org/wdb/wdb/adp/phase3_spectral/form?collection_name=HARPS}{http://archive.eso.org/wdb/wdb/adp/phase3\_spectral/form\\
?collection\_} \href{http://archive.eso.org/wdb/wdb/adp/phase3_spectral/form?collection_name=HARPS}{name=HARPS}   } In Table~\ref{tab:parameters} we list their physical parameters. As mentioned above, our hypothetical RV observations are supposed to be taken near eclipse when the primary's CCF will have zero RV. We therefore correct the CCF of 51\,Peg to be centered around zero RV throughout our computations. The RV amplitude ($K$) imposed by a 4\,d Jupiter-sized planet on an HD\,40307 is about $150\,{\rm m\,s}^{-1}$. Hence, we periodically shift the secondary's CCF according to the planet's orbital phase with an amplitude of $K=150\,{\rm m\,s}^{-1}$. We normalize the secondary CCF to the total flux of the system to properly model the brightness contrast between the two stars (see Figure~\ref{fig:CCF}). We then add the shifted CCF of the secondary and the zero-RV CCF of the primary to obtain the combined CCF (${\rm CCF}_{\rm tot}$). For different orbital phases of the planet around the secondary, we then fit a Gaussian function to ${\rm CCF}_{\rm tot}$ to approximate an observed RV measurement of the unresolved system. A similar approach was taken by \citet{2002A&A...392..215S} to study the RV signal of HD\,41004\,AB.

To estimate the ETV and EDV amplitudes of the secondary star due its planet, we utilize the analytical framework of \citet{2009MNRAS.392..181K} and \citet{2009MNRAS.396.1797K}, which has originally been developed to estimate the TTVs and TDVs of transiting planets with moons. Each individual ETV and EDV measurement in our simulations is calculated by weighing the respective amplitude with a sine function of the planetary orbital phase, ETV and EDV being shifted by $\pi/2$.

\begin{table}
\caption{Physical parameters of 51\,Peg and HD\,40307.}
\label{tab:parameters}
\begin{center}
\begin{tabular}{lccccc}
\hline
Star & Mass & Radius & $T_{\rm eff}$ & $v \sin(i)$ & $m_{\rm V}$  \\
        & $M_{\sun}$ & $R_{\sun}$ & K & ${\rm km\,s}^{-1}$ & \\
\hline
51\,Peg & 1.11 & 1.266 & 5793 & 2.8 & 5.49\\
HD\,40307 & 0.77 & 0.68 & 4977 & $<1$ & 7.17\\
  \hline
\end{tabular}
\end{center}
\end{table}

\subsubsection{An M-dwarf and a primordially transiting hot Jupiter with a Neptune-sized exomoon}

Our second test case consists of an M-dwarf and a luminous Jupiter-like transiting planet with a Neptune-mass invisible moon. The orbital periods are the same as we used for the first toy system and Hill stability is ensured. The Jupiter-like transiting planet cannot be considered as a canonical ``hot Jupiter'' because the illumination from the M-dwarf would be too weak to heat it up significantly. Hence, in order to make the planet visible in thermal emission, its intrinsic luminosity would need to be fed by primordial heat, which could be sufficient for detection during the first $\sim100$\,Myr after the formation of the system.

We model the CCF of the M-dwarf using a HARPS observed CCF of GJ\,846, an M0.5 star \citep{2002AJ....123.2002H} with a mass of approximately 0.63 solar masses ($M_{\sun}$), assuming solar age and metallicity \citep{2000ARA&A..38..337C}. Since there is no direct observation of any hot Jupiter emission spectrum, there also is no observed CCF available for our simulations. Instead, we use the CCF of 51\,Peg\,b observed by \citet{2015A&A...576A.134M} in reflected light. We consider a Gaussian function with same depth and width as the CCF of 51\,Peg\,b reported by \citet{2015A&A...576A.134M} for the transiting planet.\footnote{Note that 51\,Peg\,b is orbiting on a shorter period orbit, therefore it is much brighter in the reflected light than our test planet in a 50\,d orbit.}

Our preliminary tests showed that no moon, which can reasonably form through in-situ accretion within the gas and dust accretion disk around a young giant planet \citep[a few Mars masses at most;][]{2006Natur.441..834C,2015ApJ...806..181H}, could induce a detectable RV or ETV-EDV signature. Hence, to explore the detection limit, we assumed an extreme version of a Neptune-sized satellite around the Jovian-style transiting planet, which essentially results in a planetary binary orbiting the M-dwarf host star. The $K$ amplitude of this planetary binary, which we use to shift the planetary CCF as a function of the planet's orbital phase around the common center of mass, is around $500\,{\rm m\,s}^{-1}$.

\subsection{In-eclipse RVs and the Rossiter-McLaughlin effect}
\label{sec:RME}

As the primary star rotates, one hemisphere is moving toward the observer (and therefore subject to an RV blue-shift), while the other hemisphere is receding (and therefore red-shifted). During eclipse of a prograde binary, the secondary star first occults part of the blue-shifted hemisphere and, thus, induces an RV red-shift. During the second half of an eclipse, assuming zero transit impact parameter, the RV shift is blue. The resulting RV signal, known as the Rossiter-McLaughlin effect (RM) \citep{1924ApJ....60...15R,1924ApJ....60...22M}, has been used to determine the projected stellar rotation velocity ($v \sin i$) and the angle between the sky-projections of the stellar spin axis and the orbital plane of eclipsing binaries, and recently has been used for the transiting exoplanets in almost 100 cases as of today.\footnote{See the Holt-Rossiter-McLaughlin Encyclopaedia at\\ \href{http://www2.mps.mpg.de/homes/heller/}{www2.mps.mpg.de/homes/heller}.}

We suppose that the presence of an S-type planet around the secondary star could create an additional RV component in the RM signal because the light coming from the secondary star will show different RV shifts during subsequent eclipses (see Figure~\ref{fig:RVETVEDV_concept}). To test our hypothesis, we use the publicly available SOAP-T software \citep{2013A&A...549A..35O}\footnote{Available at \href{http://www.astro.up.pt/resources/soap-t/}{www.astro.up.pt/resources/soap-t}.}, which can simulate a rotating star being occulted by a transiting foreground object. SOAP-T generates both the photometric lightcurve and the RV signal and it can also deliver the time-resolved CCFs of the system.  

We assume that the primary star rotates with the rotation period of 24 days, which is close to the Sun's value. We simulate the eclipse of the secondary and generate the CCFs of the combined system for three orbital phases of the secondary star around its common center of mass with its S-type planet. One orbital phase simply assumes a zero RV shift of the secondary, corresponding to an orbital geometry where the secondary and its planet are both on the observer's line of sight. The second scenario assumes that the secondary is moving toward the observer with its $K$ amplitude (see Section~\ref{sec:feasibility}) during eclipse and so the secondary's CCF is maximally blue-shifted in this eclipse simulation. We then estimate the observed RV by fitting a Gaussian to all those CCFs, and obtain the RM signal that is contaminated by a redshifted secondary. Finally, the third scenario assumes a maximum redshift of the secondary during eclipse.

\begin{figure}
\includegraphics[width=1.0 \linewidth , height=5.7cm]{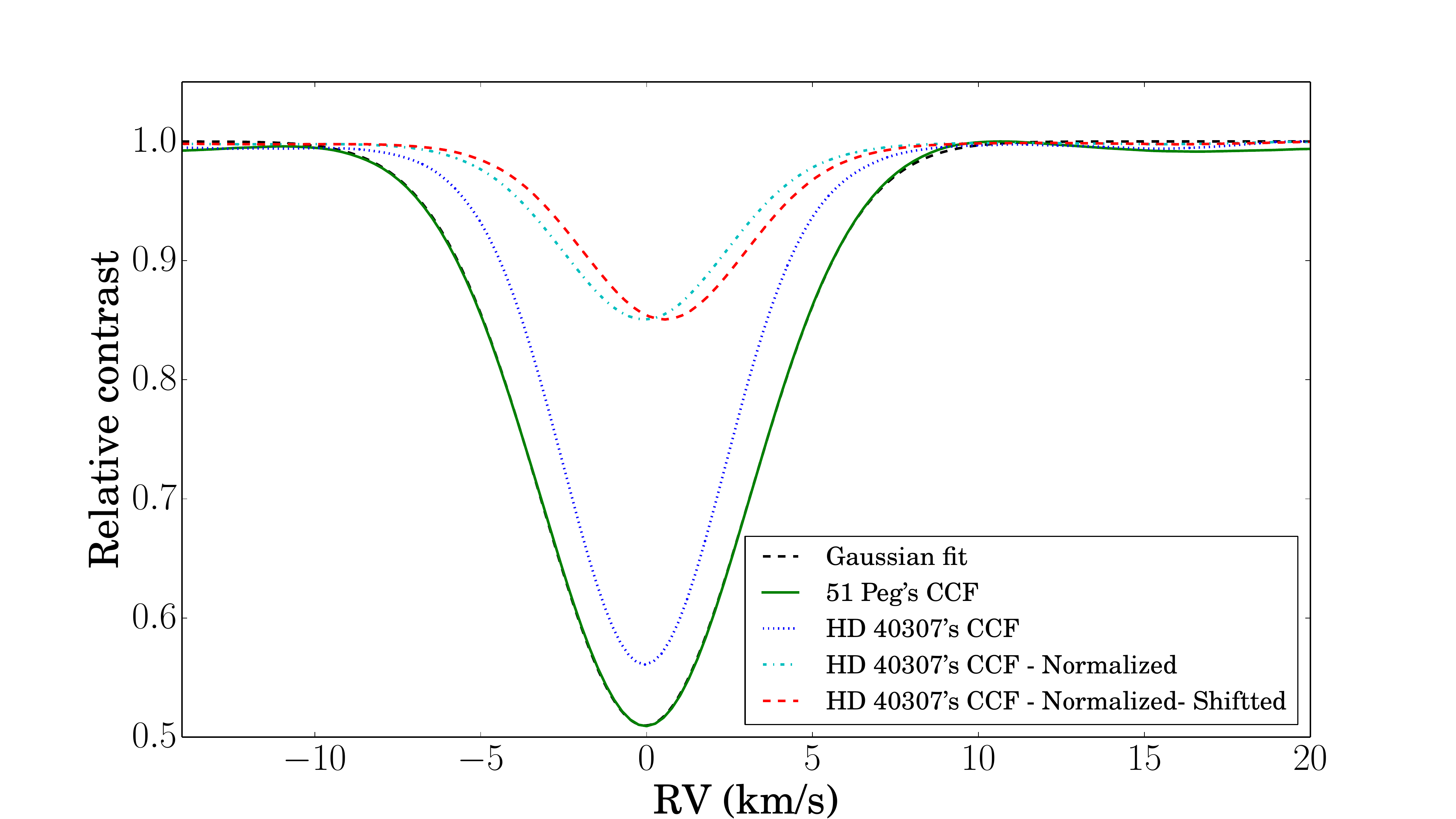}
 \caption{The solid green line represents the CCF of 51\,Peg as observed with HARPS and normalized to its continuum. The black dashed line (used in Section~\ref{sec:impact}) shows the best Gaussian fit to the CCF of 51\,Peg. The dotted blue line shows the CCF of HD\,40307 obtained with HARPS, and normalized to its continuum. The dashed dotted cyan line shows the CCF of HD\,40307 at zero RV and normalized to the total flux of system. The dashed red line illustrates the same CCF of HD\,40307 but shifted by the maximum RV induced by our S-type test planet ($K=150\,{\rm m\,s}^{-1}$).}
\label{fig:CCF}
\end{figure}

\begin{figure*}
  \centering
  \includegraphics[width=0.492\textwidth]{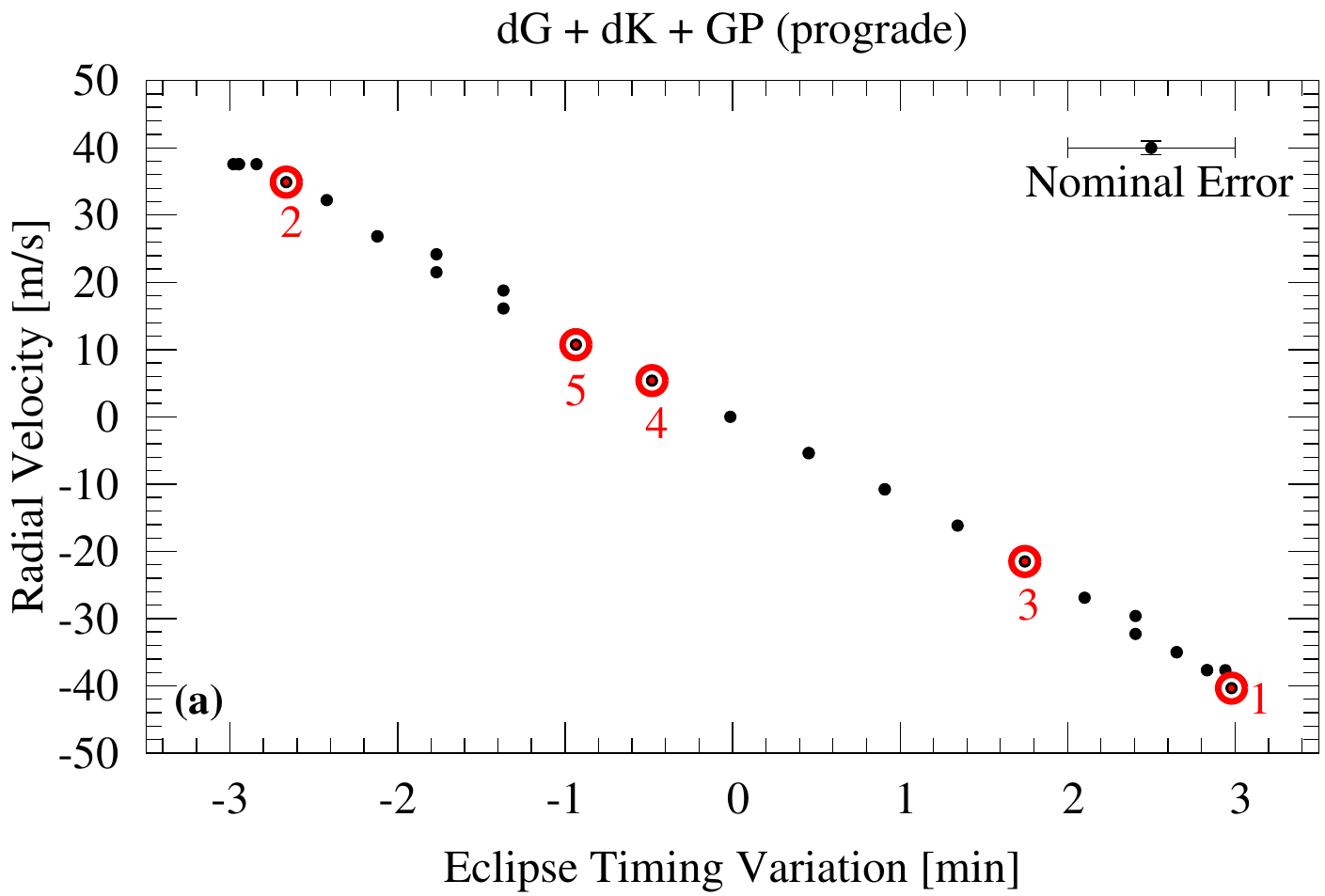}
  \includegraphics[width=0.492\textwidth]{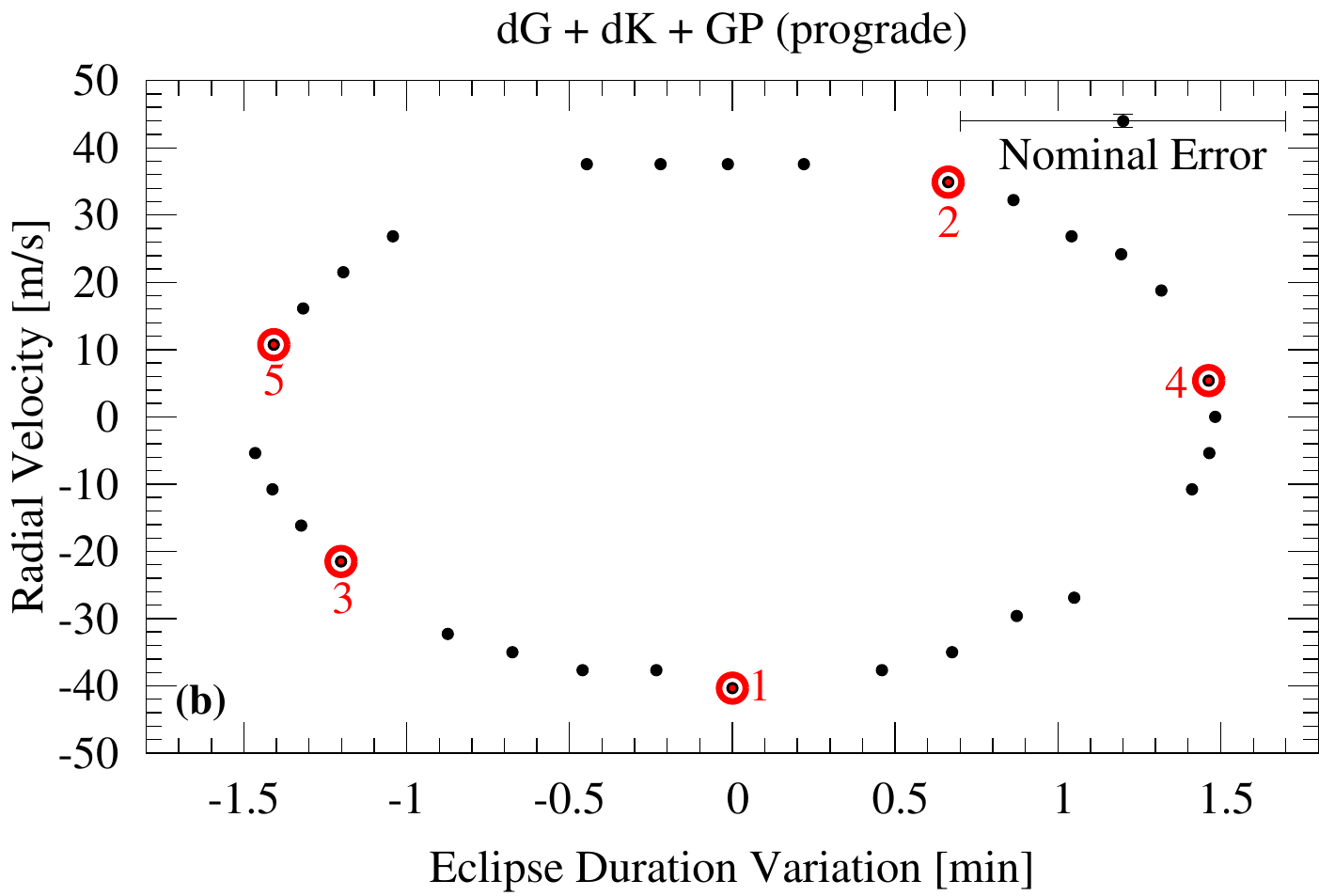}\vspace{0.5cm}
  \includegraphics[width=0.492\textwidth]{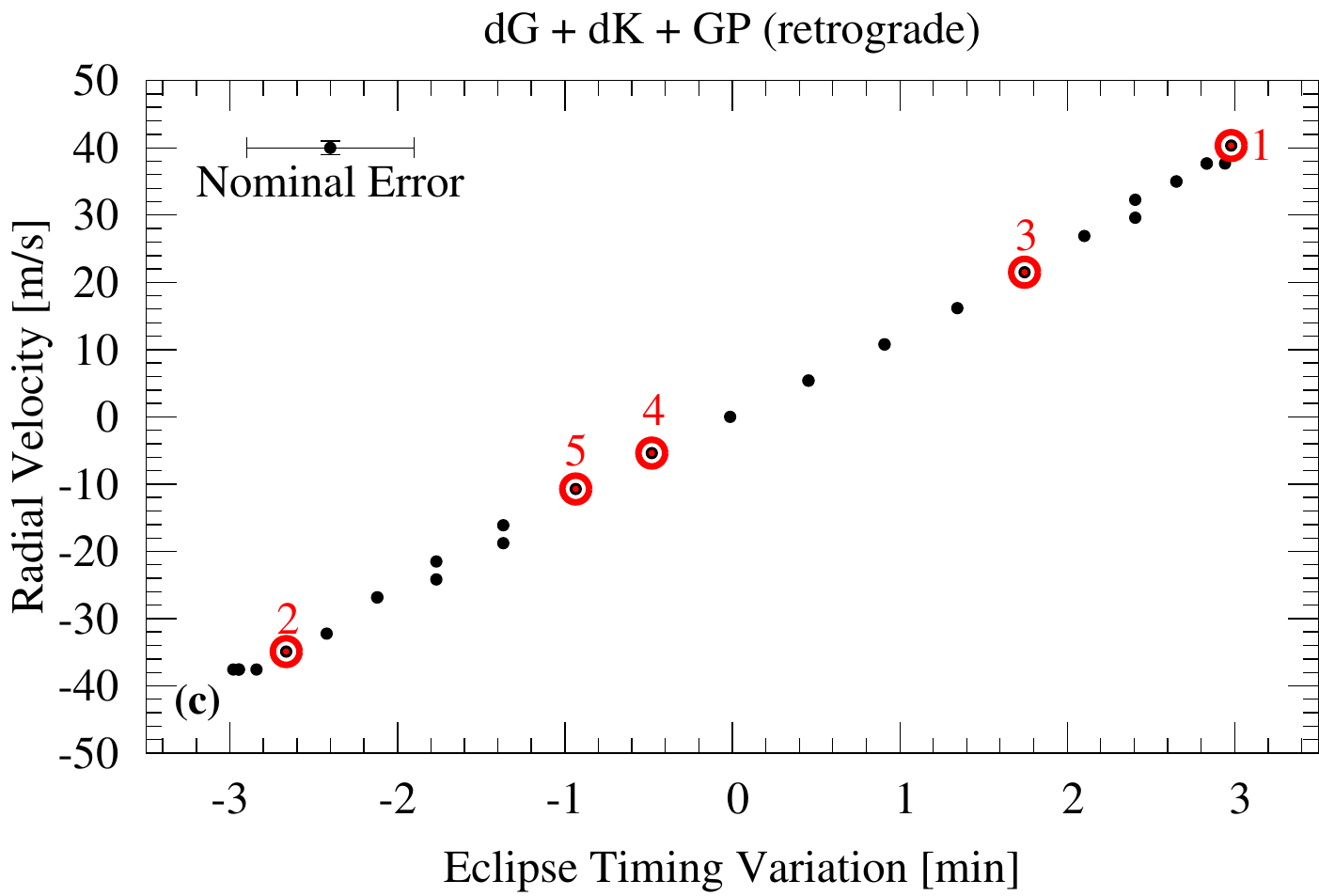}
  \includegraphics[width=0.492\textwidth]{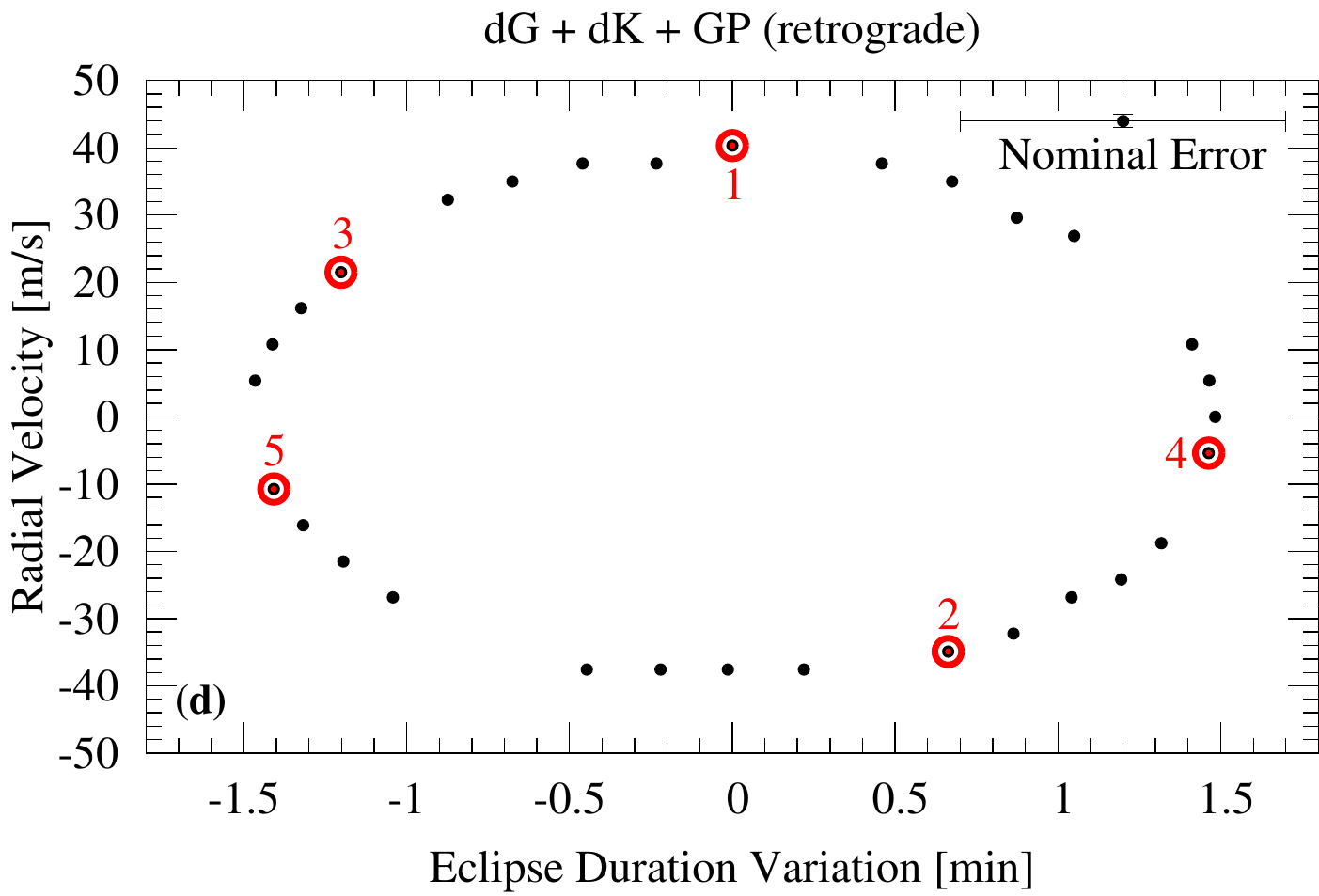}
\caption{Correlation of the RVs, ETVs, and EDVs of the secondary K-dwarf star due to a giant planet (GP) companion, both orbiting a dG primary star. In each panel, the sequence of the first five measurements is labeled with red numbers. Nominal error bars of $\pm1\,{\rm m\,s}^{-1}$ in RV and 30\,s in both ETV and EDV are shown in each panel. Panels \textbf{(a)}-\textbf{(b)} assume a prograde sense of orbital motion of the dK-GP system. Panels \textbf{(c)}-\textbf{(d)} assume a retrograde sense of orbital motion.}
\label{fig:RVETVEDV_obs}
\end{figure*}

\section{Results}

In the following, we focus on our first test case of a Sun-like host with a K-dwarf companion and a Jovian S-type planet (Section~\ref{sec:toy1}). Our second test case of an M-dwarf with a planetary binary is considered in a separate Section~\ref{sec:toy2_results}.

\subsection{RV correlations with ETVs and EDVs}

We first explore the correlation between the secondary's RVs, as measured directly from the combined dG-dK CCF (${\rm CCF}_{\rm tot}$), with their corresponding ETV and EDV observations. Figure~\ref{fig:RVETVEDV_obs} shows the RV-ETV and RV-EDV patterns for both a prograde (upper panels) and a retrograde (lower panels) S-type planet. Nominal error bars of 30\,s \citep[\textit{Kepler}'s median timing precision;][]{2013arXiv1309.1176W} for ETVs and EDVs as well as $1\,{\rm m\,s}^{-1}$ for the RV simulations (obtainable with HARPS) are shown in each panel. The sequence of the first five measurements is indicated with red symbols and numbers. As a reading example, consider measurement ``1'' in the upper row. A maximum ETV observation of $\sim+3$\, min in panel (a)\footnote{Sequences of both ETVs and EDVs always relate to an average eclipse period (for ETVs) or average eclipse duration (for EDVs). Hence, the first eclipse observation would, by definition, have no ETV or EDV. Only once multiple eclipses have been observed will the ETV-EDV measurements settle along the figures shown in Figure~\ref{fig:RVETVEDV_obs} and measurement ``1'' will turn out to have an ETV of about $+3$\,min as in this example.} means that the passage of the secondary star in front of the primary is maximally delayed. With regards to Figure~\ref{fig:RVETVEDV_concept}, the secondary has its maximum deflection to the left (see configuration labelled ``1'') so that the prograde planet is maximally deflected to the right. The prograde motion of the planet then implies that the planet is receding from the observer while the star is approaching and, hence, the latter is subject to a maximal RV blue shift of about $-40\,{\rm m\,s}^{-1}$ in Figure~\ref{fig:RVETVEDV_obs}a. Regarding the EDV component of ``1'' in Figure~\ref{fig:RVETVEDV_obs}b, it must be zero as the secondary's motion around the local secondary-planet center of mass has no tangential velocity component with respect to the observer's line of sight.

One intriguing result of these simulations is that the slopes of the RV-ETV figures for the prograde (panel a) and the retrograde planet (panel c) indeed exhibit different algebraic signs. The prograde scenario shows a negative slope, the retrograde case a positive slope, as qualitatively predicted in Section~\ref{sec:principles}. Also note that the sequence of simultaneous RV-EDV measurements in the prograde case (panel b) is a horizontally mirrored version of the retrograde case (panel d). The main advantage of our method is that by obtaining a small number of RV observations we will be able to detect the S-type planet in the system, distinguish between the prograde and retrograde orbit.

In Appendix~\ref{sec:appendix}, we also present the correlation between the Bisector Inverse Slope
(BIS) \citep{2001A&A...379..279Q} and the FWHM of the CCFs and the ETV.

\subsection{Planetary mass estimation}

\begin{figure*}
\includegraphics[width=0.75\linewidth]{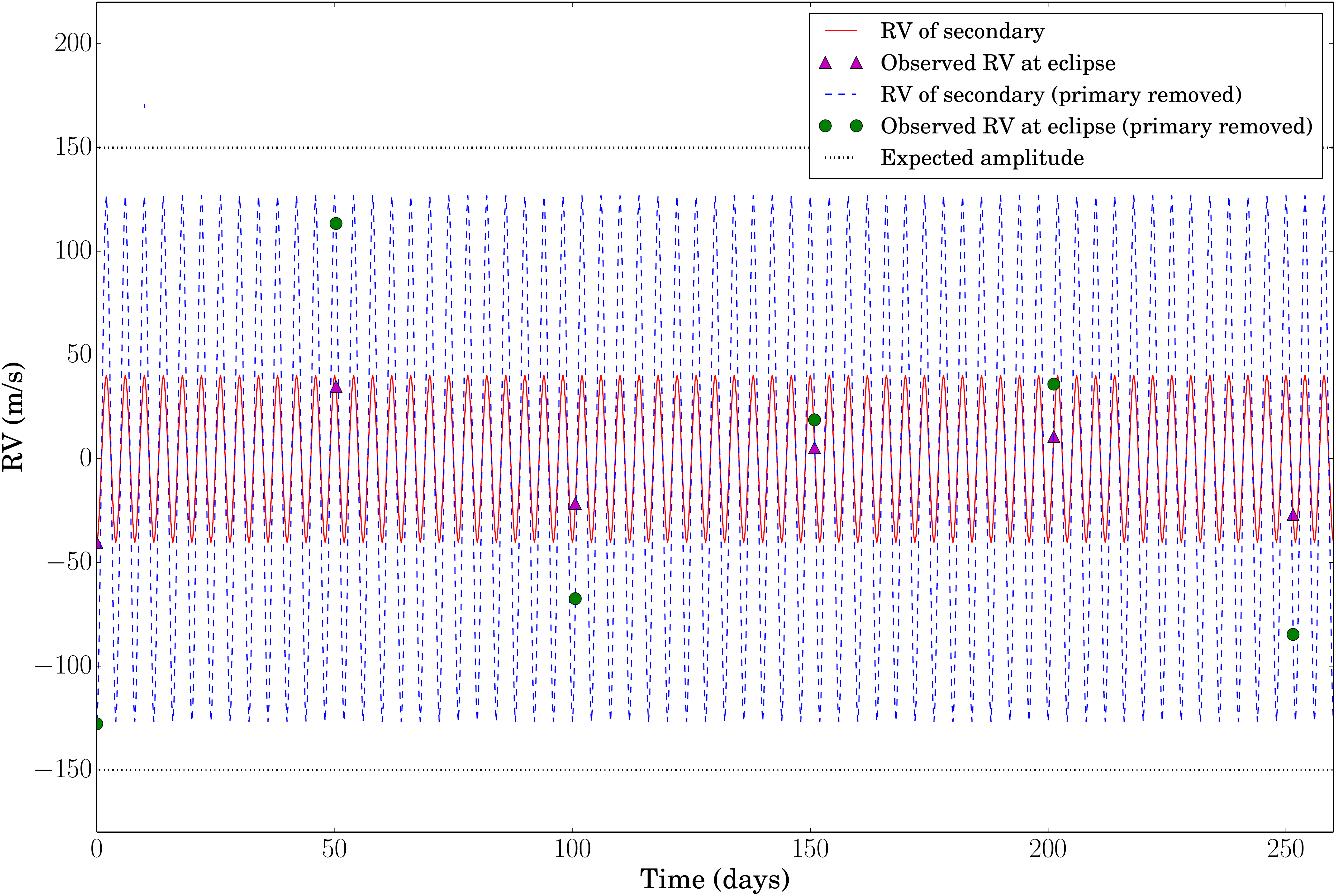}
 \caption{The red solid line represents the RV signal of the unresolved system derived by fitting a Gaussian profile to ${\rm CCF}_{\rm tot}$. Observed RVs near eclipses are represented with the magenta triangles. The dashed blue line displays the RV of the system after removing the contribution of the primary. Green circles show observed RVs near eclipses. The black dotted line at $K=150\,{\rm m\,s}^{-1}$ presents the actual amplitude of RV signal due to S-type planet. A nominal error bar of $\pm\,1\,{\rm m\,s}^{-1}$ is indicated in the upper left corner.}
  \label{fig:orbit}
\end{figure*}

By fitting the Gaussian profile to ${\rm CCF}_{\rm tot}$, we find that the RV variation of the secondary due to its Jupiter-sized S-type planet is strongly diluted by the primary star. We detect an RV amplitude of about $40\,{\rm m\,s}^{-1}$. Hence, the corresponding mass of the planet would be strongly underestimated. Our simulated RV observations are shown Figure~\ref{fig:orbit}.

To derive a more accurate measurement of planet's mass, we remove the contribution of primary light from ${\rm CCF}_{\rm tot}$. We first fit a Gaussian function to the CCF of the primary's star and then use this fit as a template representing the CCF of primary (as shown in Figure \ref{fig:CCF} in dashed black line). We then remove this template Gaussian from all simulated ${\rm CCF}_{\rm tot}$ measurements, which in reality would represent the observed CCF of the unresolved binary system.

We then fit a Gaussian function to all measurements in our simulated CCF time series, from which the primary's contribution has been removed, to estimate the RVs of the secondary due its S-type planet. We retrieve the RV variation amplitude induced by the S-type planet on the secondary star that is $85\,\%\,\pm1\,\%$ of the actual value (see Figure~\ref{fig:orbit}), which translates into an underestimation of the minimum planetary mass by $15\,\%\,\pm1\,\%$.

\subsection{Distortions of the Rossiter-McLaughlin effect}
\label{sec:impact}

While RV-ETV and RV-EDV correlations require the RV measurements to be taken either slightly before or after eclipse in order to minimize effects of the eclipsing star on the spectrum of the primary in the background (Section~\ref{sec:principles}), we now consider the actual in-eclipse RVs of the system as described in Section~\ref{sec:RME}.

\begin{figure}
\includegraphics[width=1.0\linewidth]{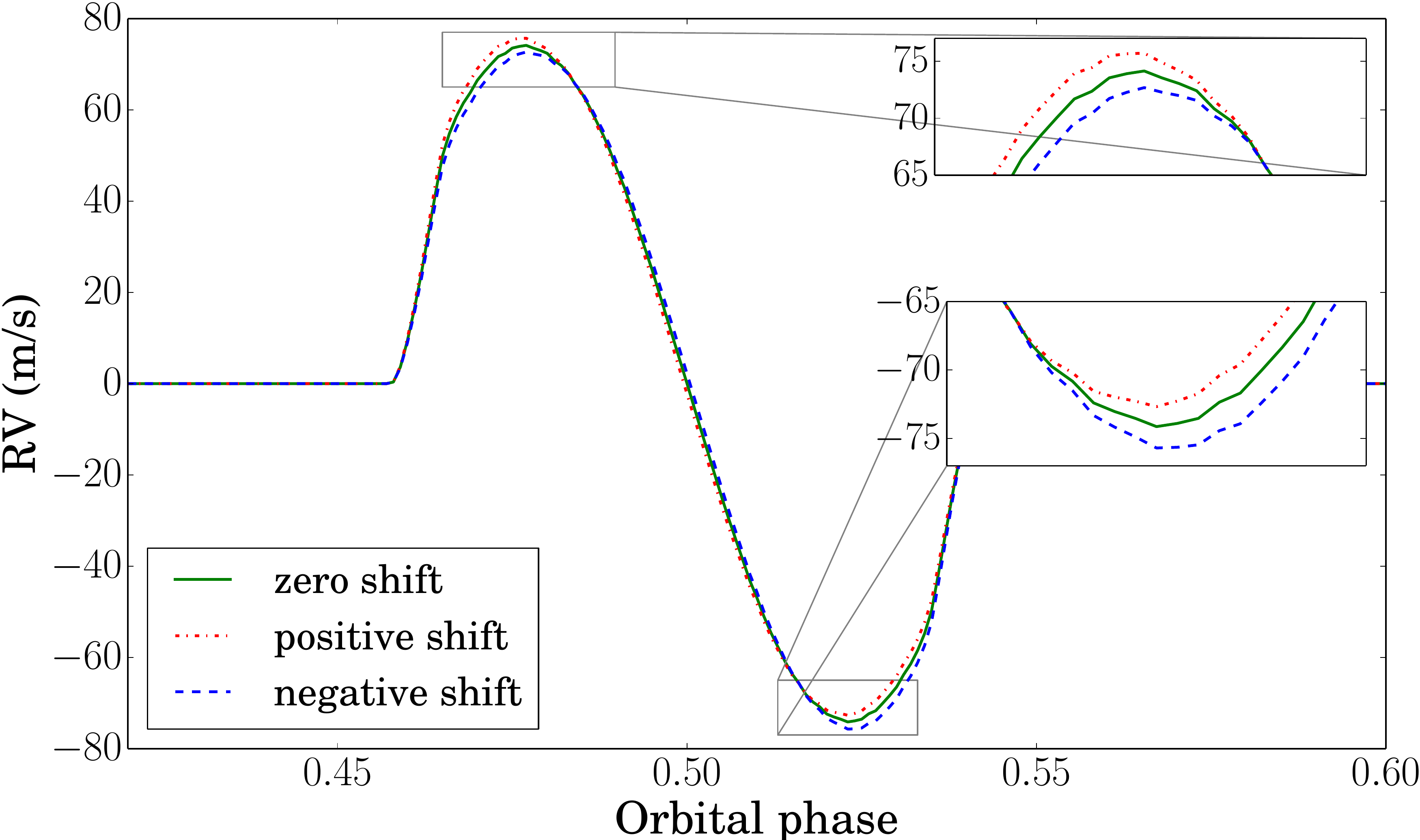}
 \caption{The green solid line shows the RM effect in the combined spectrum of a dG-dK binary with the K-dwarf star passing in front of the dG star. The dashed blue line refers to the same system, but now the K-dwarf star is subject to a maximum blue shift due to a Jupiter-sized S-type planet in a 4\,d orbit. The dashed-dotted red line refers to the reverse motion, where the secondary is moving away from the observer.}
  \label{RM-STYPE}
\end{figure}

In Figure~\ref{RM-STYPE} we show the impact of the S-type planet on the observed RM as derived from the total CCF of the dG-dK binary. Although the additional effect is only a few percent ($\sim1.5\,{\rm m\,s}^{-1}$) of the total RM amplitude ($\sim75\,{\rm m\,s}^{-1}$), it could be detectable with current high-precision spectrographs. If not taken into account properly, the asymmetry in the RM contributions could lead to inaccurate estimations of the spin-orbit angle of the eclipsing binary. But if the variation in the RM amplitude during subsequent eclipses can be correlated with the presence of an S-type planet, then it could serve as an additional means of verifying the S-type planet. That being said, note that there are several other phenomena that can affect the RM signal, such as gravitational microlensing \citep{2013A&A...558A..65O}, the convective blueshift \citep{2016ApJ...819...67C}, and occultations active region on the background star \citep{2016A&A...593A..25O}.

\subsection{Detectability of extremely massive moons around Jovian planets}
\label{sec:toy2_results}

\begin{figure*}
  \centering
   \includegraphics[width=0.492\textwidth]{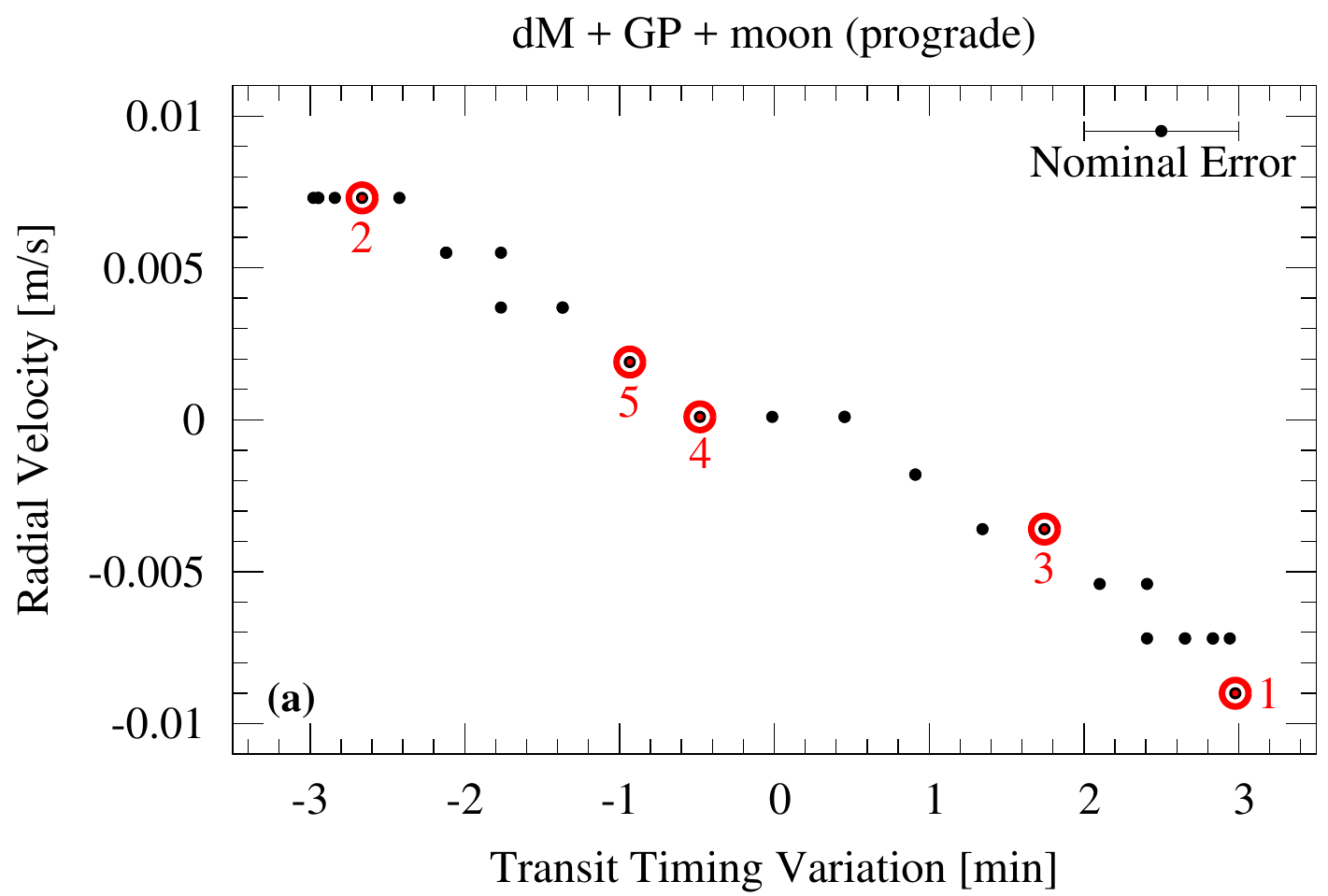}
   \includegraphics[width=0.492\textwidth]{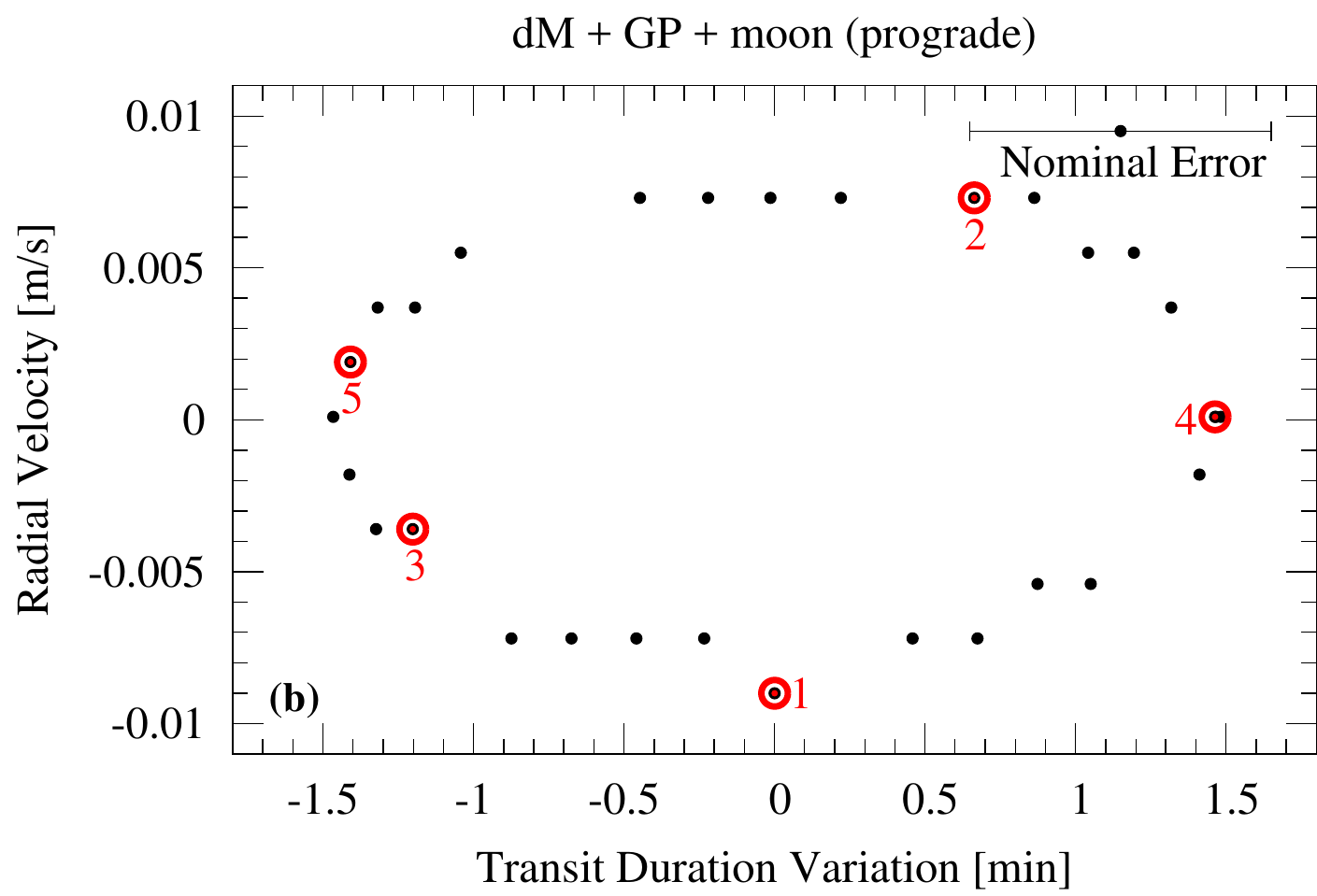}
 \caption{RVs, TTVs, TDVs of the transing planet due to a Neptune-sized moon. The host star is of spectral type M-dwarf. We assume a prograde sense of orbital motion of the exomoon system. Nominal error bars of 30\,s in both TTV and TDV are shown in each panel. Nominal error bars of RV measurements are assumed to be in the order of 10 ${\rm cm\,s}^{-1}$. }
 
 \label{fig:exomoon}
 \end{figure*}

Our simulations of an extremely massive moon around a transiting, luminous giant planet in orbit about an M-dwarf star are shown in Figure~\ref{fig:exomoon}. The predicted RV amplitude of the observations is of order of ${\rm cm\,s}^{-1}$. Note that the amplitude of the planet's actual RV motion of $500\,{\rm m\,s}^{-1}$ is much larger than the $150\,{\rm m\,s}^{-1}$ that we estimated for the K-dwarf due its S-type planet in our first test case. However, the flux ratio between a Jovian planet and its M-dwarf host star is much lower.

Hence, we find that the precision required to detect even a ridiculously massive exomoon, or essentially a binary planet, is unachievable with current RV facilities. Near future facilities such as ESPRESSO at the Very Large Telescope might be able to go down to the ${\rm cm\,s}^{-1}$ accuracy level in favorable cases \citep{2010SPIE.7735E..0FP}, so future detections of highly massive exomoons or binary planets might be possible in principle, but will certainly be challenging.

\section{Conclusion}

We present a novel theoretical method to detect and verify S-type planets in stellar eclipsing binaries by correlating the RVs of the secondary star with its ETVs and EDVs. We test the applicability of our method by performing realistic simulations and find that it can be used to detect a short-period S-type planet (e.g. a hot Jupiter) around a K-dwarf or lower-mass secondary star in a moderately wide orbit ($\sim~50$\,d) around a Sun-like primary star, e.g. using \textit{Kepler} photometry and HARPS RV measurements. We also find that the RV-ETV diagram can be used to distinguish between prograde and retrograde S-type orbits. The sense of orbital motion is a key tracer of planet formation and migration, so the RV-ETV correlation identified in this paper could be very useful in studying the origin of close-in planets in binaries. 


We show that the removal of the primary's CCF from the combined CCF of the unresolved stellar binary can yield realistic estimates of the planetary mass through RV measurements. With the ETV-EDV relation offering a methodologically independent measurement, we find that combined RV-ETV-EDV observations offer a means to both detect and confirm/validate S-type planets at the same time. ETV-EDV measurements also deliver the planet's orbital semi-major axis around the secondary star. In this paper, we propose that RV observations be taken near eclipse in order to correlate them with ETVs and EDVs. After about a dozen eclipses, or if additional RV measurements could be taken far from eclipse, it could be possible to securely identify the planet's orbital period around the secondary star. And if the secondary's mass can be estimated from its spectrum and using stellar classification schemes (e.g. stellar evolution models), then RVs could also yield an independent measurement of the planet's semi-major axis, which needs to be in agreement with the value derived from the ETV-EDV data.

Various physical phenomena can mimic RV-ETV and RV-EDV correlations, such as the stellar activity. Transiting planets crossing stellar active regions, for instance, can cause TTVs and TDVs \citep{2013A&A...556A..19O}. It is also well-known that active regions on a rotating star affect the CCF, and thus produce RV variations, even if the planet does not transit \citep{2001A&A...379..279Q}. We therefore presume that it is plausible that stellar activity produces some kind of correlation between RVs and ETVs or EDVs, although it might be very different from the patterns we predict for S-type planets.

Eclipsing binaries from \textit{Kepler} are usually faint with typical \textit{Kepler} magnitudes $11~{\lesssim}~m_{\rm K}~{\lesssim}~15$ \citep{2016MNRAS.455.4136B}. RV accuracy of $\sim1\,{\rm m\,s}^{-1}$ will be hard to achieve for many of these systems. The PLATO mission, scheduled for launch in 2025, will observe ten thousands of bright stars with $m_{\rm V}~<~11$ \citep{2014ExA....38..249R}, many of which will turn out to be eclipsing binaries. PLATO will therefore discover targets that allow both more accurate ETV-EDV measurements and high-accuracy ground-based RV follow-up.

\appendix

\section{Variations of the bisector span and the full width at half maximum}
\label{sec:appendix}

\begin{figure*}
  \centering
  \includegraphics[width=0.492\textwidth]{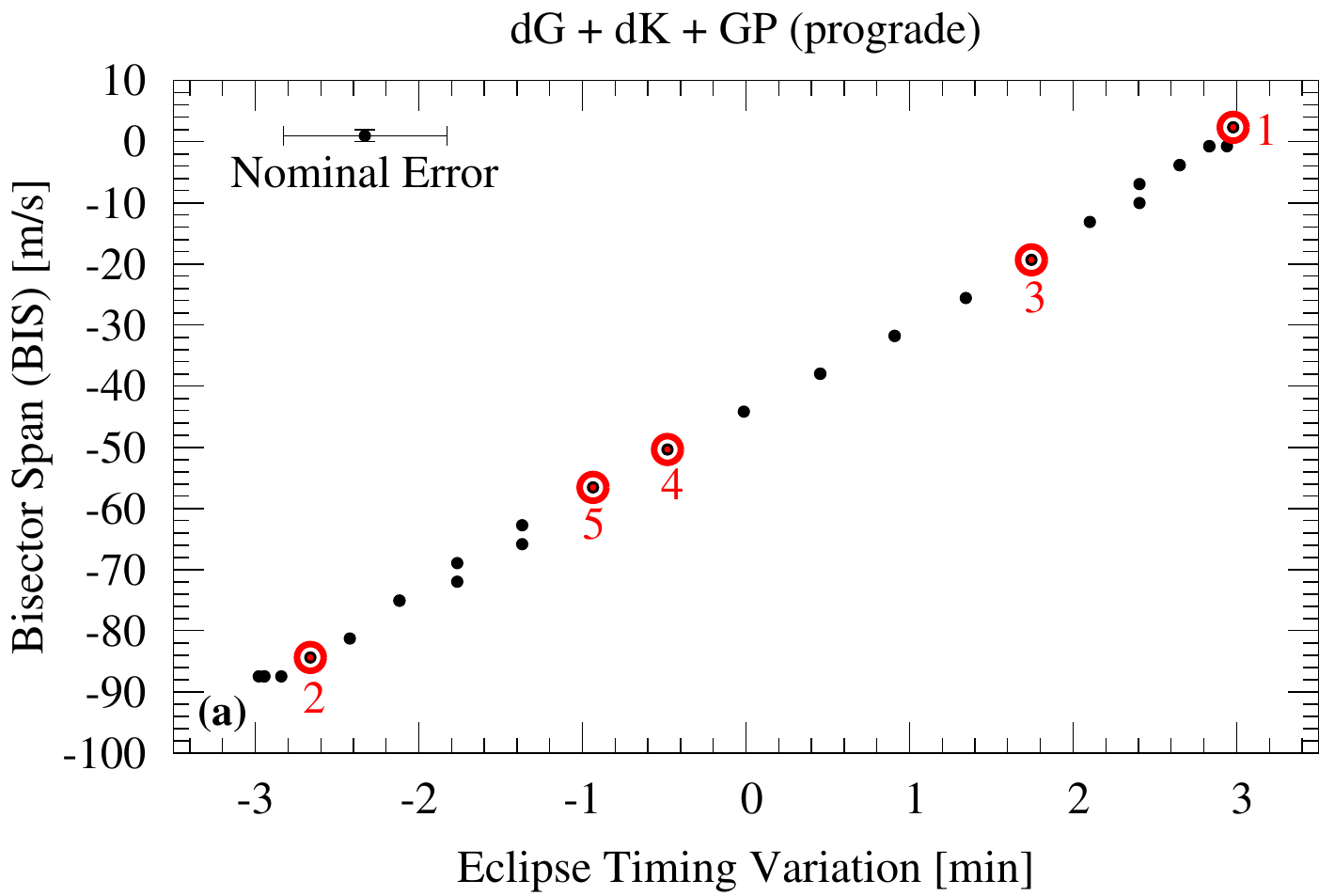}\hspace{.2cm}
  \includegraphics[width=0.49\textwidth]{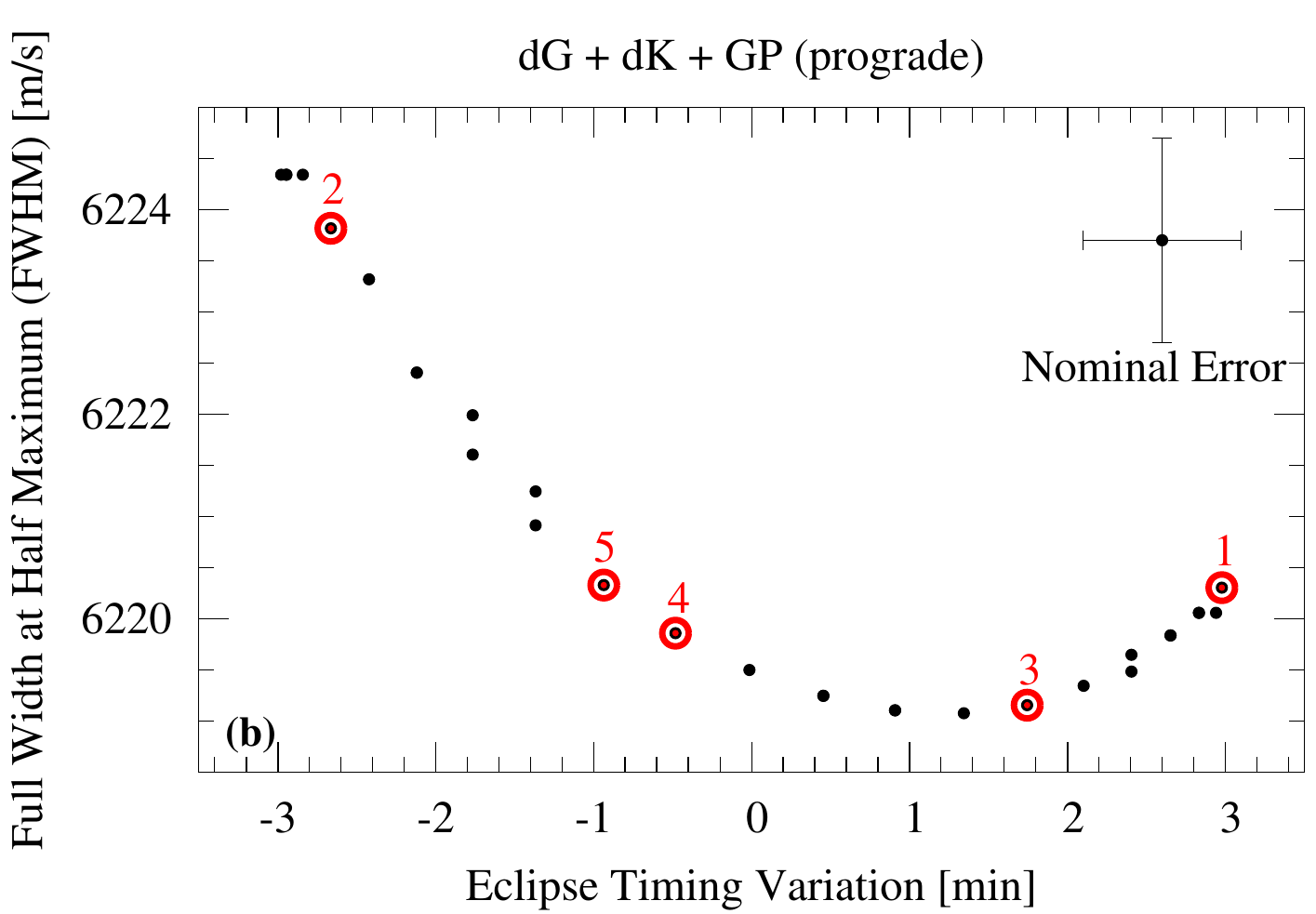}
\caption{Supplementary information to Figure~\ref{fig:RVETVEDV_obs}. Panel \textbf{(a)} shows variations of the bisector span and panel \textbf{(b)} the full width at half maximum of the CCF of the M-dwarf, both for a prograde scenario.}
\label{fig:BIS_FWHM}
\end{figure*}

As an extension of the data shown in Figure~\ref{fig:RVETVEDV_obs}, we append figures of the bisector inverse slope (BIS) \citep{2001A&A...379..279Q} and of the full width at half maximum (FWHM) of the CCFs as a function of the stellar ETVs. The data refers to the prograde scenario of a Jupiter-sized S-type planet in a 4\,d orbit around a K-dwarf star, both of which orbit a Sun-like primary star every 50\,d (see Section~\ref{sec:toy1}). Figure \ref{fig:BIS_FWHM} clearly reveals additional BIS-ETV and FWHM-ETV correlations. However, our follow-up simulations did not indicate a unique correlation between the planet's sense of orbital motion and either the BIS or the FWHM.

\section*{Acknowledgements}

\scriptsize MO acknowledges research funding from the Deutsche Forschungsgemeinschaft (DFG , German Research Foundation) - OS 508/1-1. This work made use of NASA's ADS Bibliographic Service.  We would like to
thank the anonymous referee for insightful
suggestions.




\bibliographystyle{mnras}
\bibliography{RV-ETV-EDV-MNRAS}


\label{lastpage}
\end{document}